\documentclass{PoS}

\title{Effective field theories for rooted staggered fermions}

\ShortTitle{Rooted staggered fermions}

\author{Claude Bernard\\
        Washington University, St. Louis, MO 63130, USA\\
        E-mail: \email{cb@lump.wustl.edu}}

\author{\speaker{Maarten Golterman}\\
        San Francisco State University, San Francisco, CA 94132, USA\\
        E-mail: \email{maarten@stars.sfsu.edu}}

\author{Yigal Shamir\\
  School of Physics and Astronomy,
  Raymond and Beverly Sackler Faculty of Exact Sciences\\
        Tel-Aviv University, Ramat Aviv, 69978 Israel\\
        E-mail: \email{shamir@post.tau.ac.il}}

\abstract{
We extend the construction of the Symanzik effective action to
include rooted staggered fermions, starting from a generalization of the
renormalization-group approach to rooted staggered fermions.
The Symanzik action, together with the usual construction
of a partially quenched chiral effective theory from a local,
partially quenched, fundamental theory,
 can then be used to derive
the chiral effective theory. The latter reproduces rooted staggered chiral
perturbation theory.
}

\FullConference{The XXV International Symposium on Lattice Field Theory\\
		 July 30-4 August 2007\\
		 Regensburg, Germany}

\def\ie{{\it i.e.}}

\begin{document}

\section{Introduction}

When the lattice spacing $a$ of some discretization of QCD is small enough
(\ie, $a\Lambda_{QCD}\ll 1$), effective field theories (EFTs) such as the Symanzik
effective theory (SET) \cite{Symanzik} and chiral perturbation theory (ChPT)
\cite{reviews} can be used to account for lattice artifacts through a systematic
expansion in $a\Lambda_{QCD}$.  A key assumption for this to work is that the
underlying lattice theory is local.

This raises the question whether the construction of a SET and ChPT can be
extended to QCD with staggered fermions if the ``fourth-root trick'' is used.
This trick amounts to taking the fourth-root of the staggered determinant for
each physical flavor (up, down and strange).   Each staggered quark yields a
four-fold degeneracy in the continuum limit (there are four ``tastes'' for each
physical flavor), and taking the fourth root aims at removing this degeneracy.%
\footnote{For a review of rooted staggered fermions relevant for this work,
see Ref.~\cite{bgslatt06}. For general reviews, see Refs.~\cite{sharpelatt06,kronfeld}.}
However, at non-zero lattice spacing this formulation of lattice QCD does not
correspond to a local theory \cite{bgsloc}, and thus  the question whether
the construction of EFTs can be extended to rooted staggered QCD is raised.

Here we will argue that this can indeed be done, starting from the
renormalization-group (RG) analysis of rooted staggered fermions \cite{shamirrg}.
The intuitive idea is to start with $n_r$ replicas of each staggered flavor,
leading to the presence of the $n_r$-th power of the staggered determinant for
that flavor in the lattice path integral.  This corresponds to a local theory as long
as $n_r$ is a positive integer, and the assumption is that EFTs such as the SET and
ChPT can be constructed for such $n_r$.  One then reaches the EFT for the rooted theory by
continuing $n_r\to 1/4$, which at the lattice QCD level precisely corresponds to taking the
fourth root.  The problem is, of course, whether this continuation
can be carried out, and whether it is unique.  At the EFT level, $n_r$ dependence
arises in two ways: explicitly, through the loops calculated in the EFT, but also
implicitly, through the dependence of the coefficients of the EFT (the Symanzik
coefficients in the SET, and the low-energy constants (LECs) in the chiral theory)
on $n_r$.
We will argue that the continuation exists, and is unique,
if one works to a fixed order in the lattice spacing and in the number of loops
in the EFT.  At the ChPT level, we will show that the proper EFT is
staggered ChPT (SChPT), with the replica rule reproducing the effects of
rooting. In other words it is rooted staggered ChPT (rSChPT) \cite{rschpt}.

We note that a complementary argument for the validity of rSChPT already exists \cite{rschptcb}.
A key difference with the current work is that Ref.~\cite{rschptcb} argues
completely within the context
of chiral effective theories, starting from a case (four degenerate flavors)
where the chiral theory is known because the rooting is trivial.
Here, we start instead from the fundamental theory on the lattice and show
how  rSChPT may be derived from it, {\it via}\/ the SET.
The replica rule is given definite meaning in the fundamental theory,
so its appearance in the EFTs is completely natural.
In contrast, the replica
rule in Ref.~\cite{rschptcb} has meaning only at the chiral level.
Further, Ref.~\cite{rschptcb} uses certain plausible, but unproven, assumptions on decoupling and on
the analyticity of the expansion around positive quark mass. In particular, the decoupling
assumption leaves a small potential loophole.  While the three-flavor
chiral theory goes over, in the continuum limit, to the standard three-flavor chiral theory of QCD,
it is not guaranteed that the LECs have the same numerical values as in QCD.
 The current argument dispenses
with several of the assumptions of Ref.~\cite{rschptcb} and closes the loophole:
The continuum low-energy
constants are automatically those of QCD with the correct number of flavors.
On the other hand the current argument, based as it is on Ref.~\cite{shamirrg}, inherits
the assumptions of that work.  The key assumptions are two: For any $n_r$, integrating
out ultra-violet fermionic degrees of freedom modifies the effective gauge-field action
by local terms.  In addition, because (it is highly plausible that)
the theory is renormalizable for any $n_r$,
the perturbative scaling laws apply even though the underlying theory is non-local.  (For more discussion of these assumptions, see also Ref.~\cite{bgslatt06}.)
Both the present arguments and those of Ref.~\cite{rschptcb}
rely heavily on the validity of the standard partially quenched chiral theory \cite{bgpq}
for describing partially quenched fundamental theories that are local.
We also need to assume here that the SET exists for partially-quenched theories,
as long as the lattice theory is local.

The existence of valid EFTs at non-zero lattice spacing is very important in practice.
Even at $a=0.06$~fm, and with the use of improved staggered fermions, lattice
artifacts, such as taste splittings in hadronic multiplets, are significant, and one has
to use EFT techniques that incorporate lattice artifacts
to obtain good fits \cite{MILC}.
Note that the link established here between
the validity of rSChPT and the RG analysis of Ref.~\cite{shamirrg} turns any numerical
test of the EFT framework into a more direct
test of the RG picture of the continuum limit of QCD with rooted staggered quarks.

In the next section, we review the RG framework.   We employ that framework
in Sec.~3 to generalize the staggered theory in a way that will
turn out to be useful for our goal, and we
present our main argument.  In Sec.~4 we clarify the way one of the most important
staggered symmetries, shift symmetry, works at the EFT level, and we end with
our conclusions.  This talk gives a brief account of a more detailed article
in preparation \cite{bgseft}.

\section{The renormalization-group framework}

The RG approach of Ref.~\cite{shamirrg} starts from the staggered Dirac operator in the
``taste basis'' \cite{saclay}:
\begin{equation}
\label{taste}
D^{-1}_{taste}=\frac{1}{\alpha}+QD^{-1}_{stag}Q^\dagger\ ,
\end{equation}
where $Q$ is the gauge-covariant unitary transformation taking the staggered Dirac operator from the
one-component to the taste basis.
The only new element here is the addition of the contact term $1/\alpha$.  Contact terms
like this arise naturally when one defines a series of blocked staggered Dirac operators
through gaussian RG blocking, and one may indeed
implement Eq.~(\ref{taste}) as a ``zeroth'' blocking step  in which no thinning out of
degrees of freedom occurs \cite{shamirrg}.

After carrying out $n$ blocking steps, the partition function may be written as
\begin{equation}
\label{Z}
Z(n_r)=\int{\cal D}{\cal U}\prod_{k=0}^n{\cal D}{\cal V}^{(k)}
{\bf B}_n\left(n_r;{\cal  U},{\cal V}^{(k)}\right){\rm Det}^{n_r}\left(D_{taste,n}\right)\ ,
\end{equation}
with $n_r$ is the number of replicas.  For simplicity we consider the case of
mass-degenerate flavors, with a common quark mass $m$.  For now we will
take $n_r$ to be a positive integer.  $D_{taste,n}$ is the staggered Dirac operator
after $n$ hypercubic blocking steps, and ${\cal V}^{(k)}$ are the blocked gauge fields.  As made
explicit in Eq.~(\ref{Z}), we postpone the integration over gauge fields in order to
keep the action bilinear in the fermions, or, equivalently, to express the fermionic
part of the partition function as a determinant.  We will not need the detailed form
of ${\bf B}_n$, but only note that, under the assumptions of Ref.~\cite{shamirrg},
it represents a local Boltzmann weight on the coarse lattice reached after $n$
blocking steps.  The coarse lattice spacing $a_c$ is related to the original
lattice spacing $a_f=a$ by $a_c=2^{n+1}a_f$ (on the taste basis the lattice
spacing is $2a_f$), and we choose $a_f$ and $n$
such that $a_c\Lambda_{QCD}\ll 1$.

So far, we have been following a standard RG set up.  We now define
\begin{equation}
\label{Dinv}
{\tilde{D}}_{inv,n}=\frac{1}{4}{\rm tr}_{taste}(D_{taste,n})\ ,
\end{equation}
and use this to split $D_{taste,n}$ into a taste-invariant and a taste-breaking piece:
\begin{equation}
\label{split}
D_{taste,n}={\tilde{D}}_{inv,n}
\otimes{\bf 1}+\Delta_n\ ,
\end{equation}
where ${\bf 1}$ is the $4\times 4$ unit matrix in taste space.
It can be shown that, because of the contact term $1/\alpha$ in Eq.~(\ref{taste}),
${\tilde D}_{inv,n}$ has no fermion doublers \cite{bgsloc,shamirfree}.
We now assume that the taste-breaking part scales as
\begin{equation}
\label{scaling}
||a_c\Delta_n||\sim\frac{a_f}{a_c}\ ,
\end{equation}
which expresses the assumption that taste violations scale to zero when $a_f\to 0$
like a dimension-five irrelevant operator,
in the RG language.  Note that, with $n_r$ integer, the theory is local, and there
is no reason to doubt the existence of the continuum limit.

For $n$ large enough, \ie, $||a_c\Delta_n||$ small enough, we may expand the theory
defined by $D_{taste,n}$ around the theory defined by $D_{inv,n}={\tilde{D}}_{inv,n}
\otimes{\bf 1}$.  If $\Delta_n$ scales as assumed, both theories will have
the same continuum limit,\footnote{The continuum limit is defined by taking $n\to\infty$,
keeping $a_c$ fixed.}  \ie, the continuum limits of the theory with
$n_r$ staggered fermions, and the theory with $n_s=4n_r$ taste-singlet fermions
with Dirac operator ${\tilde{D}}_{inv,n}$ are the same.

The claim of Ref.~\cite{shamirrg} is that this conclusion should also hold for the case
$n_r=n_s/4$, with $n_s$ any positive integer, not necessarily a multiple of four.
Of course, $n_r=1/4$ corresponds to a theory with one staggered quark with the
fourth root of the staggered determinant.  The key point
is that the scaling of $\Delta_n$ should also hold on the ensemble of gauge
configurations generated with $n_r=n_s/4$.  Next, we will use
this as a starting point for the construction of EFTs for rooted staggered QCD.

\section{Generalized theory}

We now generalize the theory defined in Eq.~(\ref{Z}) by replacing (for details,
see Ref.~\cite{bgseft})
\begin{equation}
\label{generalize}
{\rm Det}^{n_r}\left(D_{taste,n}\right)
\rightarrow
{\rm Det}^{n_s}\left({\tilde{D}}_{inv,n}\right)
\frac{{\rm Det}^{n_r}\left({\tilde{D}}_{inv,n}
\otimes{\bf 1}+t\Delta_n\right)}
{{\rm Det}^{n_r}\left({\tilde{D}}_{inv,n}
\otimes{\bf 1}\right)}\ .
\end{equation}
This is a theory with $n_r$ ``generalized'' staggered fermions (generalized because of the
insertion of the interpolation parameter $t$), $n_s$ taste-singlet fermions, and $4n_r$
taste-singlet ghosts (\ie, quarks with bosonic statistics).  For $t=1$ and $n_s=4n_r$ the
generalized theory reduces to the staggered theory, Eq.~(\ref{Z}), while for $t=0$ it is
a local ``reweighted'' theory of $n_s$ taste-singlet fermions.

According to our basic assumption, that the SET and ChPT
exist for any local discretization of QCD, we may thus assume that the SET, and ChPT,
can be constructed along the usual lines for any positive integers $n_r$ and $n_s$, and any $t$.
We will in particular assume that the SET still exists for any fixed integer $n_s$,
not necessarily equal to $4n_r$ (while still keeping $n_r$ integer), for any value of
$t$.  In other words, we will assume that the SET exists for partially-quenched theories
\cite{bgpq}, as long as the lattice theory is local.  Since ${\tilde D}_{inv,n}$ is defined on the
coarse lattice, the theory defined by the replacement (\ref{generalize}) depends on both the
fine and coarse lattice spacings $a_f$ and $a_c$, and we will think about the SET as an
expansion in $a_f$, with coefficients that depend on $a_c$.

The key observation is that the determinant ratio in Eq.~(\ref{generalize}) can be expanded
in $t$, if $\Delta_n$ is small enough (\ie, if the number of RG steps $n$ is large enough):
\begin{equation}
\label{expand}
\frac{{\rm Det}^{n_{\scriptstyle r}}\left({\tilde{D}}_{inv,n}
\otimes{\bf 1}+t\Delta_n\right)}
{{\rm Det}^{n_r}\left({\tilde{D}}_{inv,n}
\otimes{\bf 1}\right)}=
{\rm exp}\left[n_r{\rm Tr}\log\left(
1+t({\tilde{D}}_{inv,n}^{-1}\otimes{\bf 1})
\Delta_n\right)\right]\ .
\end{equation}
Since $\Delta_n$ scales like $a_f$, and the trace over taste indices of $\Delta_n$ vanishes,
it is easy to see that this expansion is really a combined expansion in $t$, $n_r$, and $a_f$,
with the power of $n_r$ smaller than the power of $t$, and the power of $t$ smaller than or
equal to
that of $a_f$ in each term.  This implies that when we expand any lattice correlation function
to a given fixed order in $a_f$, any such correlation function is polynomial in $n_r$.\footnote{For
a careful treatment of external legs, as opposed to the loops arising from the determinant,
see Ref.~\cite{bgseft}.}  In the SET that reproduces these correlation functions to that same
order in $a_f$, the $n_r$ dependence comes from Symanzik coefficients as well as from loops.
To a given order in the loop expansion, the dependence coming from loops is polynomial
in $n_r$, and we thus conclude that the Symanzik coefficients have to be polynomial in $n_r$ as well.
Having established that the SET, including its coefficients as well as
all of its correlation functions, depends polynomially
on $n_r$ to any order in $a_f$, it follows that the analytic continuation
to non-integer $n_r$ exists, and is unique.  We may now set $n_r=n_s/4$ and $t=1$,
obtaining the correct SET for QCD with $n_s$ (degenerate) rooted staggered fermions.
This is our main result.

We now summarize a number of important comments and consequences.
For $t\ne 1$ the SET is complicated, with dependence on both lattice spacings, $a_f$ and
$a_c$.  However, we are ultimately interested only in correlation functions at $t=1$, and usually
with only staggered fermions on the external legs.  In that case, all staggered symmetries (in
particular shift symmetry and $U(1)_\epsilon$ symmetry) apply, and it follows that the
form of the SET is precisely that of Ref.~\cite{ls}, to order $a_f^2$.  (Ref.~\cite{ls} only dealt with the
case $n_r=1$, but it follows from the discussion in Ref.~\cite{rschpt} that an analogous form can be
written down for integer $n_r\ne 1$ as well.)
If we now set $n_r=n_s/4$,
the underlying lattice theory is just lattice QCD with $n_s$ rooted staggered quarks,
and there is only one lattice spacing $a_f$.  (While correlation functions will
still depend on $a_f$ and $a_c$, physical quantities can only depend on $a_f$.)

The $t$-expansion is an expansion in ${\tilde D}_{inv,n}^{-1}\Delta_n$.  This
is similar to
the expansion in $D_{cont}^{-1}(D_{latt}-D_{cont})\sim ap$
underlying the usual construction of the SET, with $p$ the
typical momentum on the external legs.  Here $\Delta_n\sim a_fp^2$ and
${\tilde D}_{inv,n}^{-1}\sim 1/p$, where the momentum $p$ is restricted to be smaller than the
cutoff $1/a_c$, because the theory defined by ${\tilde D}_{inv,n}$ lives on the
coarse lattice.  It follows that the $t$-expansion is an expansion in powers of $a_f/a_c$.
The theory has no $1/a_f$ divergences, consistent with the
fact that the continuum limit is the same for all $t$.  (At $t=1$ there are no $1/a_f$ divergences
because of staggered symmetries \cite{gs}, while at $t\ne 1$ the theory is $O(a_f)$ away from the
$t=1$ theory.)

The transition to the chiral theory works in the same way, leading us to SChPT with the replica rule
(rSChPT)
\cite{rschpt,rschptcb} as the correct chiral theory for QCD with rooted staggered fermions.
As emphasized above, we need to assume that the chiral theory for the local, but partially
quenched, theory with integer $n_r$ is a standard (graded-symmetry) one \cite{bgpq}.

\section{How shift symmetry works at the effective level}

Shift symmetry plays an important role in the construction of any EFT for QCD with
staggered fermions, and here we briefly discuss what shift symmetry looks like at the
effective level.  Shift symmetry is generated by the transformations
\begin{equation}
\label{shift}
S_\mu\chi(x)=\zeta_\mu(x)\chi(x+\mu)\ ,\qquad\qquad
\zeta_\mu(x)=(-1)^{x_{\mu+1}+\dots+x_4}\ ,
\end{equation}
where $\chi$ is the staggered field in the one-component formulation.  From this it follows that
any representation of the group takes the form \cite{gs,girrep}
\begin{eqnarray}
\label{irrep}
&&S_\mu\to e^{ia_fp_\mu}\Xi_\mu\ ,\ \ \ \ \ (-\pi/2<a_fp_\mu<\pi/2)\ ,\\
&&\{\Xi_\mu,\Xi_\nu\}=2\delta_{\mu\nu}\ .\nonumber
\end{eqnarray}
However, any continuum EFT is invariant under continuum translations, which, for a translation
by a displacement $r$, act on continuum fields as
\begin{equation}
\label{translation}
\phi(p)\to e^{ip\cdot r}\phi(p)\ .
\end{equation}
We may now combine Eqs.~(\ref{irrep}) and (\ref{translation}) with $r$ such that $p\cdot r=-a_fp_\mu$,
and conclude that the EFT is invariant under the group $\Gamma_4$ generated by the $\Xi_\mu$.

\section{Conclusions}

We have shown that the usual construction of a Symanzik effective theory
can be extended to lattice QCD with rooted staggered fermions.  An important corollary is that
SChPT with the replica rule (rSChPT) is the correct chiral theory for this lattice discretization of
QCD.  These results follow rather straightforwardly from the RG framework
of Ref.~\cite{shamirrg}; thus an important link is established between the RG-based
argument that rooted staggered QCD has the correct continuum limit, and the
effective theory already in use for fitting results from numerical computations.
Turning this around, the success of such numerical fits constitutes
a rather direct test of the validity of the rooted theory.  This lends strong support to the
conjecture that, while the rooted theory is non-local
at $a\ne 0$, the non-local behavior is a lattice artifact
that can be understood and
investigated systematically in an expansion in the lattice spacing.

The generalized lattice theory described by Eq.~(\ref{generalize}) provides
us with separate handles on the number of physical flavors, $n_s$,
and the discretization effects of the original $n_r$ staggered fields,
which become non-local whenever $n_r$ is analytically continued to
non-integer values.  The continuum limit, in which the number of RG-blocking
steps $n$ goes to infinity at fixed coarse lattice spacing $a_c$,
is a local theory  (the coarse-lattice action is a ``perfect'' action) in the correct
universality class, that depends on $n_s$ but not on $n_r$.
The lattice theory itself
is independent of $n_r$ at $t=0$, and turning on the parameter $t$ re-introduces the $n_r$ dependence.
Correlation functions, calculated to any given order in $a_f$ and to any order
in the loop expansion in the effective theories (SET or SChPT), are polynomials
in $n_r$. We may thus analytically continue $n_r$ to $n_s/4$ and arrive at rSChPT.

The argument here is complementary to an earlier argument,
formulated within the chiral effective theory itself, in support of the validity
of rSChPT for rooted staggered QCD \cite{rschptcb}.
The main new element is the direct linkage between the Symanzik effective theory
and the underlying discretization of QCD, in its RG-blocked form.
Thus SChPT with the replica rule, derived via the Symanzik effective theory,
is now directly tied to the underlying rooted theory.  It follows that not only does
rSChPT give the correct form of the
dependence of physical quantities on the quark mass, but also that the continuum low-energy
constants are those of QCD with $n_s$ flavors.

For $n_r=n_s/4$ with $n_s$ not a multiple of four, the theory is non-local, and this non-local
behavior is reproduced by the EFT.  The way this happens is via the application of
the replica rule: one calculates a correlation function
in the EFT with $n_r$ replicas of staggered fermions,
setting $n_r=n_s/4$ at the end of the calculation.
Once $n_r$ has been continued to non-integer values,
the correlation functions of the $a_f\ne 0$ theory cannot be reproduced by
any local lagrangian.
For an instructive example of how this
works in rSChPT, we refer
to the discussion of the scalar meson two-point function to one loop in SChPT in Refs.~\cite{rschptcb,bgseft}.

\vspace{3ex}
\noindent {\bf Acknowledgments}
\vspace{3ex}

CB and MG were supported in part by the US Department of Energy.  YS was
supported by the Israel Science Foundation under grant no.~173/05.

\end{document}